# A low power DAQ system with high-speed storage for submersible buoy

Zhilei Zhang, Peng Miao, Houbing Liu, Kun Hu, Feng Li, Ge Jin

*Abstract–Submersible Buoy (SB) is an important apparatus capable of long-term, fixed-point, continuous and multi-directional measurement of acoustic signals and hydrological environment monitoring in the harsh marine environment, providing important information for hydrological environment research, marine organism research and protection. We will describe a real-time data acquisition (DAQ) system with multiple designs to meet low-power consumption and high-speed data transmission.*

## I. Introduction

The SB has long runtimes and recovery cycles, and the entire system is powered by batteries. So, the DAQ system should be as low-power as possible due to the requirement of long-term, stable underwater operation. Meanwhile, the DAQ system also demands large capacity storage because of long runtime and relatively high sampling rate. Owing to the wide distribution of the SB, the DAQ system must ensure high-speed data storage as well as high-speed upload path to the host when retrieving data in order to reduce data collection time and improve the extraction efficiency.

To deal with such situations, we design a data acquisition system, which consists of multi-channel ADCs for hydrophones signal digitization; Spartan-6 FPGA for data package, capture and compression; SD card for short-term data storage; SSD for long-term data storage along with dual Gigabit Ethernet for data upload. Preliminary test shows that the DAQ system can achieve high-speed data conversion, storage and readout at comparatively low-power.

## II. System Design

Through analyzing the system requirements, characteristics and feasibility of the SB, this paper describes the architecture of the DAQ system: the Front-End board (FEB) for data acquisition and packaging; the SD card board for data compression and short-term data storage; the SATA board for long-term data storage and high-speed uploading.

<>Manuscript received June 23, 2018. This work was supported by the National Natural Science Foundation of China under Grants 11461141010 and 11375179, and in part by "the Fundamental Research Funds for the Central Universities" under grant No. WK2360000005.
Zhilei Zhang, Peng Miao, Houbing Liu, Kun Hu, Feng Li, and Ge Jin are with State Key Laboratory of Particle Detection and Electronics, University of Science and Technology of China, Hefei, Anhui 230026, P.R. of China (phone: +86-551-63606495, e-mail: zzlei@mail.ustc.edu.cn, mpmp@mail.ustc.edu.cn, luhb@mail.ustc.edu.cn, khu@.ustc.edu.cn, phonelee@ustc.edu.cn, goldjin@ustc.edu.cn)

### A. Front End board

The schematic structure of the FEB is shown in Fig. 1. The core of the FEB consists of a Spartan-6 FPGA and eight hydrophones with 5~10K sampling rates. The watch circuit immediately sends a trigger signal to the FEB when monitoring the signal of interest, and the FEB starts to collect and package the data of each hydrophone and other sensors. Each packet composes a data frame after adding packet number, header of frame, end of frame and CRC16 check digit. Finally FEB sends data frame to the SD card board through the serial port.

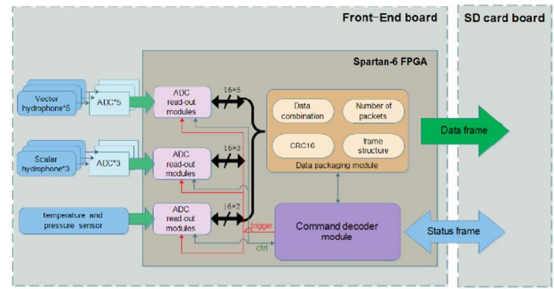

Fig.1. Schematic Block diagram of the Front-End board

### B. SD card board

The schematic block diagram of the SD card board is illustrated in Fig. 2. The SD card board compresses the frame data which is received from the FEB and write to SD card. FPGA control two SD cards based on ping-pong operation to prevent data loss. Meanwhile, either SD card which is not working is complete shut down by FPGA. When a SD card reaches storage threshold, the SATA board will be powered-on in the next data acquisition. SD mode is used to write SD card at 10MB/s for low-power consumption and UHS mode is used to read SD card data at 90MB/s for high-speed transmission with SATA board.

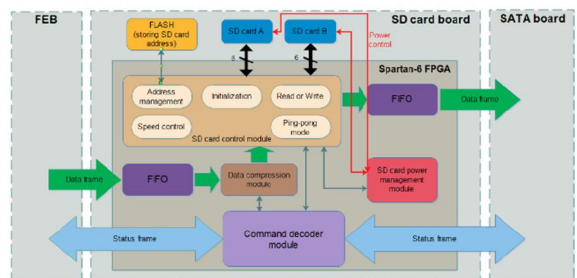

Fig.2. Schematic block diagram of the SD card board

## C. SATA board

The schematic block diagram of the SATA card board is illustrated in Fig. 3. The SATA board receive the frame data from the SD card board and write to 1TB SSD. In order to reduce power consumption, FPGA will power-off either SSD which is not at work. When the host requires data retrieve, all data in the SD cards is forcibly transferred to the SSD. Then the host read out all data from SSD through the dual Gigabit Ethernet. Despite the high power consumption of the SATA board, it runs only 0.3% of the total system time and does not affect the design goal of low-power.

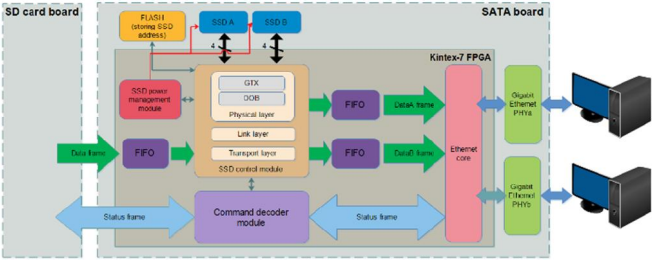

Fig.3. Schematic block diagram of the SATA board

## III. LOW-POWER DESIGN

### A. Device Selection

Use Spartan-6 FPGA for low-power consumption. (Spartan-6 FPGA can not meet the requirements of SATA boards. The Kintex-7 FPGA is chosen for high power consumption, but the running time of the SATA board is about 0.3% of the Front-End/SD card board, which does not affect the design of the low-power.)

### B. Power Management

The FEB is powered on by the watch circuit only when a target of interest is found and normally powered off to reduce power consumption. After the initialization of FEB, the SD card board is powered on by the FEB. In addition, the SD card board usually works in the ping-pong mode and the FPGA will completely shut down the idle SD card. The SATA board is powered on only when the SD card board reaches the threshold or the host computer needs to extract data . The idle SSD will be powered off by the FPGA on SATA board when the SATA board is operating properly. All digital parts of hardware are powered by DC/DC to reduce power consumption.

### C. Lossless data compression

Currently the lossless data compression encode based on FPGA are mainly RLE encode, Huffman encode, and LZW encode. By comparing the 1.5GB actual measured data from Qingdao Lake which used different lossless compression encode (shown in Fig. 5.), the compression efficiency of RLE encode is too low. The compression efficiency of LZW encode is a bit lower than Huffman encode. In general, Huffman coding is easy to accomplish, and the addition of power consumption is less than 10mW. Using lossless data compression greatly increases the effective data transmission speed, reduces system operating time, and decreases power consumption in disguise. [2][3]

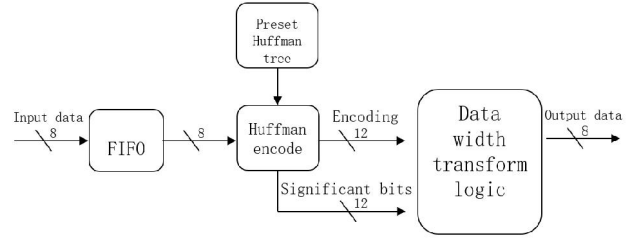

Fig.4. Schematic block diagram of lossless data compression

|  | RLE encode | Huffman encode | LZW encode |
|---|---|---|---|
| 1st test (260MB data) | 63.34% | 46.23% | 48.87% |
| 2nd test (260MB data) | 67.14% | 49.65% | 50.23% |
| 3rd test (260MB data) | 65.28% | 47.81% | 50.16% |
| 4th test (260MB data) | 63.72% | 47.42% | 49.71% |
| 5th test (260MB data) | 64.59% | 47.75% | 50.04% |
| 6th test (260MB data) | 64.67% | 46.90% | 49.95% |

Fig. 5. Data compression comparison

## IV. EXPERIMENTAL RESULTS

### A. High-speed transmission Test

In FEB FPGA, all of inputs are replaced with controllable linear-growth counters to complete data transfer processes. The received data and the transmission time of every 2048 data frames are monitored on the SD card board and the SATA board. Our tests show that the FEB generates total of 8*107 bytes, and the data received on the SD card board and the SATA board keeps increasing linearly. In other words, there are no errors in the transmission. The average speed of the SD card board and SATA board is shown in Fig. 6. In addition, each data frame has an independent packet number, and the packet number received from SATA board data has also been continuous, indicating that the data packet has never been lost. [4]

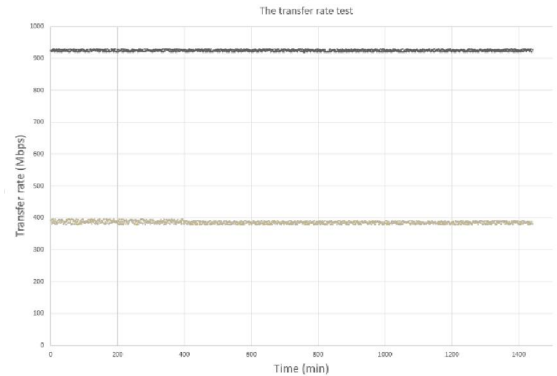

Fig. 6. Transfer rate test of SD card and gigabit ethernet

### B. Lossless data compression

The result of lossless data compression has shown in Fig. 5. , which used actual measured data from Qingdao Lake. It indicates that the power consumption is reduced a half under the same conditions, meanwhile the effective data speed is doubled and the limit of transmission speed on hardware is breakthrough.

## V. CONCLUSION

This paper realizes large capacity storage and high-speed data transmission through research and application of SD card UHS mode, SATA protocol and Gigabit. Analyze and practice from various aspects to reduce system power consumption. The prototype of the FEB, the SD card board (shown in Fig.7.), and the SATA board (shown in Fig.8.) was designed and verified for its function and performance. As a result, The DAQ system satisfies the requirement of the target.

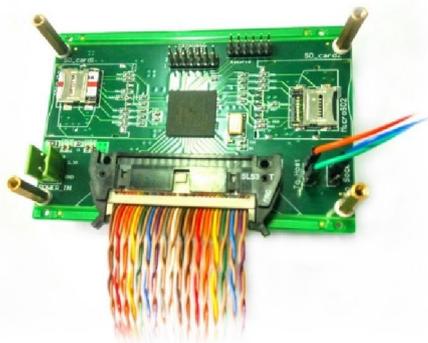

Fig.7. prototype of SD card board

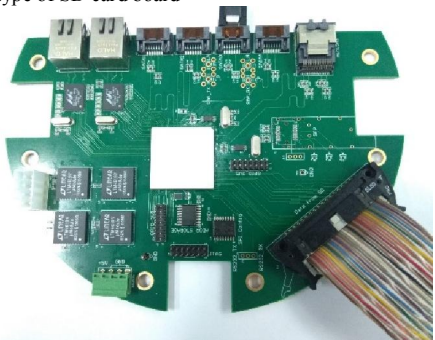

Fig.8. prototype of SATA board

REFERENCES

[1] Fan Y, "FPGA-based data acquisition system," in Signal Processing, Communications and Computing (ICSPCC), 2011 IEEE International Conference on. IEEE, 2011, pp. 1-3.
[2] Glass L, Biran G, Golander A. Compression algorithm incorporating dynamic selection of a predefined huffman dictionary: U.S. Patent 8,610,606[P]. 2013-12-17.
[3] Lekatsas H, Henkel J, Wolf W. Code compression for low power embedded system design[C]//Proceedings of the 37th Annual Design Automation Conference. ACM, 2000: 294-299. J V Hatfield, S Bell and P I Neaves. A wedge and strip particle detector based on a current-mode approach [J]. Circuits & Systems, 1995, 1:193-196.
[4] Lu H, Hu K, Wang X, et al. High speed ethernet application for the trigger electronics of the new small wheel[C]//Real Time Conference (RT), 2016 IEEE-NPSS. IEEE, 2016: 1-4.